\newcommand{\be}{\begin{eqnarray}}
\newcommand{\ee}{\end{eqnarray}}

\documentclass[twocolumn,aps,showpacs]{revtex4}
\usepackage{graphics}
\begin{document}
\title{$\rho^0$ mass in a hot hadron gas}  
\author{Alejandro Ayala}  
\affiliation{Instituto de Ciencias Nucleares, Universidad Nacional 
         Aut\'onoma de M\'exico, Apartado Postal 70-543, 
         M\'exico Distrito Federal 04510, M\'exico.}

\begin{abstract}

We study the behavior of the $\rho^0$ vector mass during the hadronic
phase of an ultra-relativistic heavy-ion collision at finite
temperature. We show that scattering with the most abundant particles
during this stage, namely, pions, kaons and nucleons, leads to an overall
temperature dependent, decrease of the $\rho^0$ intrinsic mass at
rest, compared to its value in vacuum. The main contributions arise
from $s$-channel scattering with 
pions through the formation of $a_1$ resonances as well as with
nucleons through the formation of even parity, spin 3/2 [N(1720)] and
5/2 [$\Delta$(1905)] nucleon resonances. We show that it is possible
to achieve a shift in the intrinsic $\rho^0$ mass of order $\sim - 40$
MeV, when including the contributions of all the relevant mesons and
baryons that take part in the scattering, for temperatures
between chemical and kinetic freeze-out. 

\end{abstract}


\maketitle

\section{Introduction}

The $\rho$ vector meson plays a special role in ultra-relativistic
heavy-ion collisions since its vacuum life time ($\sim 1.3$ fm) is 
shorter than the life time of the system created in the
reaction. Therefore, any changes in the properties of this meson are
expected to comprise information about the conditions of the collision
at the time when the meson decayed. This expectation is at the core of
the intensive studies, both experimental~\cite{experiments} and
theoretical~\cite{Models} regarding the $\rho$ electromagnetic decay
channels. 

The advent of large multiplicity events at RHIC energies has allowed
to also study the $\rho$ decay into pions, which is by far the most
important of its decay channels. A remarkable result 
reported by the STAR collaboration is a shift of $\sim -40$ MeV and
$\sim -70$ MeV for the peak of the invariant mass distribution of the
decay $\rho^0\rightarrow\pi^+\pi^-$ in minimum bias p + p and peripheral
Au + Au collisions, respectively, at $\sqrt{s_{NN}}=200$ GeV as
compared to the vacuum value~\cite{STAR}.  

A plausible interpretation of the above result, as regards to the
heavy-ion environment, is that one is in fact looking at the medium
induced modifications of the meson driven by its decay, regeneration
and ree-scattering within a hadronic system over a short
interval of time, namely, the last stage of the collision, between
chemical and kinetic freeze out, lasting a time interval of the order
of the life time of the meson. The fact that the resonance spectral
density can be experimentally reconstructed means that the 
hadronic system is dilute enough so that the decay products do not
suffer significant ree-scattering after being produced.

In order to describe medium induced modifications to the {\it intrinsic}
properties (mass and width) of a hadronic resonance such as $\rho$, it
is necessary to resort to effective Lagrangians representing the
interactions of the resonance with the rest of the hadronic matter in
the environment. Based on general grounds, these Lagrangians are built
to respect the basic symmetries of the 
strong interaction, among them, current conservation and parity
invariance. Thermal modifications to the $\rho$ intrinsic properties
are computed by evaluating the one-loop modification of its
self-energy. Attention is  paid to those hadrons whose rest mass is
near the threshold for $s$ or $t$ channel resonance formation and with
a sizable coupling to $\rho$ and to the most abundant particles in the
hadronic phase of the collision, namely pions/kaons and nucleons. The
phenomenological coupling constants are evaluated by comparing the
model prediction of the vacuum decay rate, into the given channel,
with the experimentally measured branching ratio.

Temperature driven modifications to intrinsic
properties of $\rho$ have been thoroughly worked out from interactions
with mesons~\cite{Rapp} when considering only scattering in the
$s$-channel. The case of interactions with baryons has
mainly been studied in connection with changes caused by dense
nuclear matter effects (see however Refs.~\cite{Rapp3}) and by means
of non-relativistic approximations 
for the interaction Lagrangians~\cite{Teodorescu, Friman, Peters}. In
a recent work~\cite{acm} interactions of $\rho$ with baryon resonances
have been considered in a relativistic framework. When including also
the contributions from scattering off various meson resonances, it has
been shown that the intrinsic $\rho$ mass at rest decreases with
increasing temperature.

The purpose of this work is to give full account of the thermal
modifications to the mass of the $\rho^0$ meson when considering its
scattering off pions/kaons and nucleons in a thermalized hadronic
medium, such as the one that is expected to be produced during the
dilute, almost baryon free, last stage of an ultra-relativistic
heavy-ion collision. In particular we show how a systematic
thermal field-theoretical calculation of the real part of the one-loop
$\rho$ self-energy yields the total contribution, both form $s$- and
$t$-channel resonance exchanges, to the thermal mass of $\rho$ at
rest. We find that $\rho$ scattering off pions through the exchange of
pions themselves significantly contributes to an increase of the
$\rho$ thermal mass. However, when considering also the contribution
from scattering off pions through the exchange of pseudo-vector
resonances, in particular $a_1$, and off nucleons through the exchange
of various baryonic resonances, in particular N(1720) and $\Delta
(1905)$, the overall result is a decrease of the thermal $\rho$ mass
at rest.  

To describe all the interactions of $\rho$ we work within a full
relativistic formalism, making use of the generalized Rarita-Schwinger
propagators for fermion fields with spin higher than 1/2. We work in
the imaginary-time formulation of thermal field theory to compute the
one-loop $\rho$ self-energy $\Pi_{\mu\nu}$, when interacting with the
relevant hadrons. 

The work is organized as follows: In Sec.~\ref{secII} we review
essentials of the computation of the real part of the one-loop thermal
self-energy of a particle in a scalar model, 
separating explicitly its $s$- and $t$-channel contributions. In
Sec.~\ref{secIII} we compute the contribution to the $\rho^0$ thermal
mass from interactions with nucleons and baryon resonances. We also
compute the coupling constants used in the calculation comparing the
theoretical expression for the branching ratio into the $\rho$ nucleon
channel with the experimentally measured value. In Sec.~\ref{secIV}
we compute the contribution to the $\rho^0$ thermal mass from
interactions with pions and kaons and various other meson
resonances. In Sec.~\ref{secV} put together the above contributions to
compute the behavior of the $\rho^0$ mass at rest with
temperature. Finally, we summarize our results and conclude in
Sec.~\ref{secVI}. 

\section{Thermal self-energy in a scalar theory}\label{secII}

Recall that the self-energy $\Pi$ is related to the intrinsic $\rho$
properties by ${\mbox{Im}}\ \Pi = -M_\rho\Gamma^{\mbox{\tiny{tot}}}$,
$M_\rho =\sqrt{m_\rho^2+{\mbox{Re}}\ \Pi}$,
where $m_\rho$ is the mass of $\rho$ in vacuum, $M_\rho$ and 
$\Gamma^{\mbox{\tiny{tot}}}$ are the (temperature and/or density
dependent) intrinsic mass and total decay width
of the $\rho$ meson, respectively and ${\mbox{Re}}\ \Pi$ and
${\mbox{Im}}\ \Pi$ represent the real and imaginary parts of $\Pi$,
respectively. 

The real part of the one-loop $\rho$ self-energy receives
contributions stemming from the various particles (fermions or bosons)
in the loop interacting with $\rho$. Depending on the Lorentz nature
of the interacting particles, the tensor structure of the self-energy
can become very cumbersome and difficult to keep track of. Nevertheless,
the analytic structure of the induced changes to the mass of $\rho$ at
finite temperature can already be understood in a simpler scalar
theory. Let us therefore here look at such case.

\begin{figure}[th] 
\hspace{-2cm}
{\centering
\resizebox*{0.15\textwidth}
{0.1\textheight}{\includegraphics{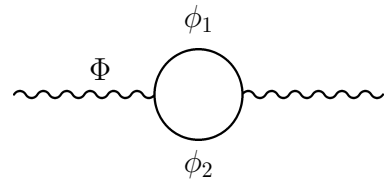}}
\par}
\caption{One-loop self-energy Feynman diagram representing the
interaction of a scalar particle $\Phi$ with two other scalar
particles $\phi_1$ and $\phi_2$.} 
\label{fig1}
\end{figure}

Consider the one-loop Feynman diagram depicted in Fig.~\ref{fig1}
representing the self energy of a scalar field $\Phi$ interacting with
other two scalar fields $\phi_1$ and $\phi_2$ through the interaction
Lagrangian 
\be
   {\mathcal {L}}_I=-{\mbox g}\Phi\phi_1\phi_2\, ,
   \label{Lintexamp}
\ee
where g is the coupling constant with dimensions of energy. Field
$\Phi$ has mass $Q^2=q_0^2-q^2=M^2$ whereas fields $\phi_1$ and 
$\phi_2$ have masses $m_1$ and $m_2$, respectively. Without lose of
generality, let us take $m_2 > m_1$.

In the imaginary-time formalism of thermal field theory, the
self-energy diagram of Fig.~\ref{fig1} can be expressed, after
performing the sum over the Matsubara frequencies, as~\cite{LeBellac}
\be
   \Pi (i\omega, q)&=& -{\mbox g}^2\int 
   \frac{d^3k}{(2\pi )^3}\frac{1}{4E_1E_2}
   \Big\{\left[1+n(E_1)+n(E_2)\right]\nonumber\\ 
   &&\left(\frac{1}{i\omega -E_1
   -E_2} -\frac{1}{i\omega +E_1+E_2}\right)\nonumber\\ 
   &-&
   \left[n(E_1) - n(E_2)\right]\nonumber\\
   &&\left(\frac{1}{i\omega -E_1
   +E_2} -\frac{1}{i\omega +E_1-E_2}\right)\Big\}\, ,
   \label{selfscalar}
\ee
where $E_1=\sqrt{k^2+m_1^2}$,
$E_2=\sqrt{({\mathbf{q}}-{\mathbf{k}})^2+m_2^2}$ and
\be
   n(x)=\frac{1}{e^{\beta x} - 1}\, ,
   \label{BE}
\ee
is the Bose-Einstein distribution with $1/\beta =T$ being the
temperature. 

The retarded self-energy is obtained by means of the analytic
continuation $i\omega\rightarrow q_0 + i\epsilon$. In this manner, the
real and imaginary parts of the retarded self-energy are given by
\be
   \!\!\!\!\!{\mbox{Re}}\ 
   \Pi (q_0,q)&\!\!\!\!\!\!\!\!\!\!=\!\!\!\!\!\!\!\!\!\!&
   -{\mbox g}^2{\mathcal{P}}\int
   \frac{d^3k}{(2\pi )^3}\frac{1}{4E_1E_2}\nonumber\\
   &\!\!\!\!\!\!\!\!\!\!\!\!\!\!\!\!\!\!\!\!&
   \Big\{\left[1+n(E_1)+n(E_2)\right]\nonumber\\ 
   &\!\!\!\!\!\!\!\!\!\!\!\!\!\!\!\!\!\!\!\!&\left(\frac{1}{q_0 -E_1
   -E_2} -\frac{1}{q_0 +E_1+E_2}\right)\nonumber\\ 
   &\!\!\!\!\!\!\!\!\!\!-\!\!\!\!\!\!\!\!\!\!&
   \left[n(E_1) - n(E_2)\right]\nonumber\\
   &\!\!\!\!\!\!\!\!\!\!\!\!\!\!\!\!\!\!\!\!&\left(\frac{1}{q_0 -E_1
   +E_2} -\frac{1}{q_0 +E_1-E_2}\right)\Big\}\, ,
   \nonumber\\
   \!\!\!\!\!{\mbox{Im}}\ \Pi (q_0,q)&=&\pi {\mbox g}^2\int
   \frac{d^3k}{(2\pi )^3}\frac{1}{4E_1E_2}\nonumber\\
   &\!\!\!\!\!\!\!\!\!\!\!\!\!\!\!\!\!\!\!\!&
   \Big\{\left[1+n(E_1)+n(E_2)\right]\nonumber\\ 
   &\!\!\!\!\!\!\!\!\!\!\!\!\!\!\!\!\!\!\!\!&
   \left(\delta (q_0-E_1-E_2) - \delta(q_0+E_1+E_2)\right)\nonumber\\ 
   &\!\!\!\!\!\!\!\!\!\!-\!\!\!\!\!\!\!\!\!\!&
   \left[n(E_1) - n(E_2)\right]\nonumber\\
   &\!\!\!\!\!\!\!\!\!\!\!\!\!\!\!\!\!\!\!\!&
   \left(\delta (q_0-E_1+E_2) - \delta(q_0+E_1-E_2)\right)\!\!\!\Big\}
   \label{realandim}
\ee
where ${\mathcal{P}}$ represents the Cauchy principal value.
Notice that the first and second of Eqs.~(\ref{realandim}) are related
by a dispersion relation without subtractions.

Let us concentrate in the first of Eqs.~(\ref{realandim}) and consider
the limit ${\mathbf{q}}=0$, that is, the situation when the particle
$\Phi$ is at rest. Furthermore, let us look only at the temperature
dependent terms. We can then write
\be
   {\mbox{Re}}\ \Pi (q_0,q)&\rightarrow&-{\mbox g}^2{\mathcal{P}}\int
   \frac{d^3k}{(2\pi )^3}\frac{1}{4E_1E_2}\nonumber\\
   &&\Big\{ n(E_1)\left[\frac{1}{q_0-E_1-E_2} - \frac{1}{q_0-E_1+E_2}
   \right.\nonumber\\
   &+&\left.\frac{1}{q_0+E_1-E_2} - \frac{1}{q_0+E_1+E_2}\right]\nonumber\\
   &+& n(E_2)\left[\frac{1}{q_0-E_2-E_1} - \frac{1}{q_0-E_2+E_1}
   \right.\nonumber\\
   &+&\left.\frac{1}{q_0+E_2-E_1} - \frac{1}{q_0+E_2+E_1}\right]\Big\}\, ,
   \label{limq0}
\ee
where $E_i=\sqrt{k^2+m_i^2}$, ($i=1,2$). $E_1$ and $E_2$ are related
by $E_2^2=E_1^2+m_2^2-m_1^2$. Notice that the coefficient of $n(E_2)$
in Eq.~(\ref{limq0}) is obtained from the coefficient of $n(E_1)$ by
the exchange $E_2\leftrightarrow E_1$. 

\begin{figure}[t!] 
\hspace{-1.5cm}
{\centering
\resizebox*{0.36\textwidth}
{0.18\textheight}{\includegraphics{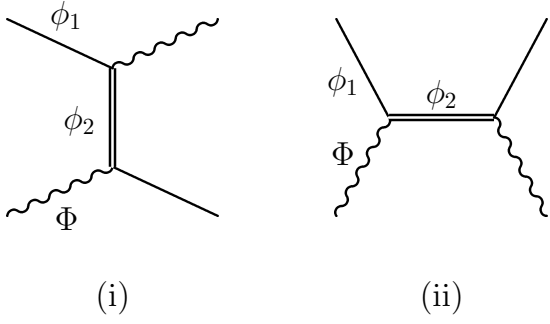}}
\par}
\caption{Two of the Feynman diagrams obtained by computing the real
part of the one-loop self-energy of $\Phi$ representing (i)
$t$-channel and (ii) $s$-channel scattering of particle $\Phi$ off
particle $\phi_1$ with the exchange of particle $\phi_2$.} 
\label{fig2}
\end{figure}

For the case at hand, namely $m_2 > m_1$, the terms proportional to
$n(E_2)$ are suppressed with respect to those proportional to
$n(E_1)$, thus, for the ease of the discussion let us for the time
being ignore the former. After the change of integration variable
$k\rightarrow E_1=\sqrt{k^2+m_1^2}$ in Eq.~(\ref{limq0}) one gets
\be
   {\mbox{Re}}\ \Pi (q_0,q=0)&=&-
   \frac{{\mbox g}^2}{(2\pi )^2}\frac{1}{2q_0}{\mathcal{P}}\int_{m_1}^\infty
   dE_1\sqrt{E_1^2-m_1^2}\nonumber\\
   &&n(E_1)\Big\{\frac{-1}{E_1 - 
   \left[\frac{q_0^2-(m_2^2-m_1^2)}{2q_0}\right]}
   \nonumber\\
   &+&
   \frac{1}{E_1 + 
   \left[\frac{q_0^2-(m_2^2-m_1^2)}{2q_0}\right]}\Big\}\, .
   \label{simple}
\ee 
For definitiveness, take $q_0 >0$. Also define
\be
   a\equiv\frac{q_0^2-(m_2^2-m_1^2)}{2q_0}\, .
   \label{eqa}
\ee
The integrand in Eq.~(\ref{eqa}) is singular for $q_0=\pm a$ when
$a\geq m_1$. We therefore have two instances for the integrand to
become singular:

\noindent
(i) $a>0$, this means that $q_0>\sqrt{m_2^2-m_1^2}$ and thus
there are singularities for $q_0\geq m_1+m_2$.

\noindent
(ii) $a<0$, this means that $0<q_0<\sqrt{m_2^2-m_1^2}$ and thus
that there are singularities for $0\leq q_0\leq m_2-m_1$. 

\begin{figure}[t!] 
{\centering
\resizebox*{0.4\textwidth}
{0.2\textheight}{\includegraphics{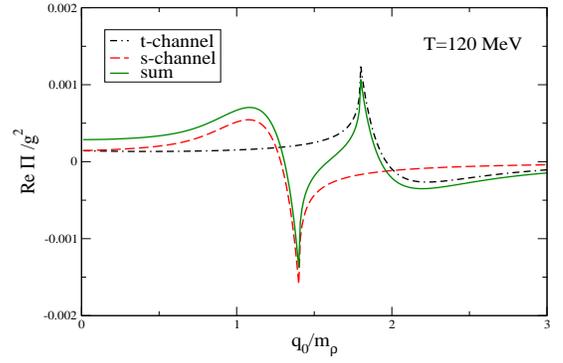}}
\par}
\caption{Real part of the one-loop self-energy for particle $\Phi$
scaled to the square of the dimension-full coupling constant g in
the scalar model showing the $t$- and $s$-channel contributions,
separate and combined. For this example we have taken $m_1=0.2
m_\rho$, $m_2=1.6 m_\rho$ and $T=120$ MeV. Notice that the $s$-channel
contribution is maximum in magnitude at $M=q_0=m_2-m_1$ whereas the
$t$-channel contribution is maximum in magnitude at
$M=q_0=m_2+m_1$. For other values of $q_0$, these contributions are of
similar magnitude showing the importance to account for both of them
in the computation of $\Pi$.}
\label{fig3}
\end{figure}

The term in Eq.~(\ref{simple}) with singularity for $a>0$ (the first
term) corresponds to the probability associated 
to $t$-channel scattering of particle $\Phi$ off particle $\phi_1$
with the interchange of particle $\phi_2$ as depicted in
Fig.~\ref{fig2}(i). Similarly, the term in Eq.~(\ref{simple}) with
singularity for $a<0$ (the second term) corresponds
to $s$-channel scattering of particle $\Phi$ off particle $\phi_1$
with the interchange of particle $\phi_2$ as depicted in
Fig.~\ref{fig2}(ii). These are the two possibilities that one obtains
when cutting the intermediate line associated with particle $\phi_1$
in the self energy diagram of Fig.~\ref{fig1}, as corresponds to the
calculation of the real part of $\Pi$.

Figure~\ref{fig3} shows the contributions to ${\mbox{Re}}\ \Pi
(q_0,q=0)$ corresponding to $t$- and $s$-channel scattering,
separate and combined. Notice that both contributions are of the same
magnitude except when $q_0\simeq m_1+m_2$ where the
$t$-channel dominates or when $q_0\simeq m_2-m_1$ where the
$s$-channel dominates. At these values of $q_0$ the integral has cusps
but it is otherwise finite since it corresponds to the Cauchy principal
value.  

The above calculation outlines the basic features of the one-loop real
part of the self-energy of a given particle in interaction with other
two as a function of its mass at rest. In the following sections we
will carry out similar considerations for the Lagrangians
corresponding to interactions of $\rho$ with several hadrons at finite
temperature, where details concerning the strength of the specific
interaction will differ but otherwise the above
general features will remain. 

\begin{table}[t!]
\vspace{0.4cm}
\begin{tabular}{|cccccc|}  \hline\hline
\multicolumn{1}{|c}{$R$}  &
\multicolumn{1}{c}{$J^P$} &
\multicolumn{1}{c}{$\Gamma^{\mbox{\tiny{vac}}}_{R\rightarrow\rho N} 
                   {\mbox{(MeV)}}$} &
\multicolumn{1}{c}{$\Gamma^{\mbox{\tiny{vac}}}_{\mbox{\tiny{tot}}} 
                   {\mbox{(MeV)}}$} &
\multicolumn{1}{c}{$f_{J^P}$} &
\multicolumn{1}{c|}{$f_{J^P}^{\mbox{\tiny{N--R}}}$} \\ \hline 
$N$(1520)      & $3/2^-$ & $25 $ & 120 &  $9.7$ &   $7$   \\   
$N$(1720)      & $3/2^+$ & $100$ & 150 &   $7$  & $7.8$   \\   
$\Delta$(1700) & $3/2^-$ & $120$ & 300 & $4.4$  &  $5$    \\
$\Delta$(1905) & $5/2^+$ & $210$ & 350 & $12.9$ & $12.2$  \\ \hline
\end{tabular}
\label{tab1}
\caption{List of baryon resonances included in the calculation. The
  last column corresponds to the values of the coupling
  constants for a non-relativistic (N-R) calculation as computed in
  Ref.~\cite{Peters}.}  
\end{table}

\section{Interactions of $\rho$ with nucleons and baryon
  resonances}\label{secIII} 

\subsection{Self-energy}

A look at the review of particle physics~\cite{Hagiwara} reveals the
existence of four baryon resonances with rest masses near the sum of
the rest masses of $\rho$ and nucleon ($N$), where, as we have seen in
Sec.~\ref{secII}, we could expect an important contribution to
the thermal modification of the $\rho$ mass. These have
also sizable decay rates into the $\rho-N$ channel. The resonances are
N(1520), N(1720), $\Delta$(1700) and $\Delta$(1905). Table~I shows
their quantum numbers and branching ratios into the $\rho-N$ channel.

The interaction Lagrangians ${\mathcal {L}}$ are given by
\be
   {\mathcal {L}} = \left\{ 
   \begin{array}{lr}
      \frac{f_{3/2^-}}
      {m_\rho}{\mathcal {F}}_{\rho N3/2^-}
      \bar{\psi}^\mu\gamma^\nu\psi F_{\mu\nu} &
      \left(\ J^P=\frac{3}{2}^-\right) \\
      \frac{f_{3/2^+}}
      {m_\rho}{\mathcal {F}}_{\rho N3/2^+}
      \bar{\psi}^\mu\gamma^5\gamma^\nu\psi F_{\mu\nu} &
      \left(\ J^P=\frac{3}{2}^+\right) \\
      \frac{f_{5/2^+}}
      {m_\rho^2}{\mathcal {F}}_{\rho N5/2^+}
      \bar{\psi}^{\mu\nu}\gamma^5\gamma^\lambda
      \psi\partial_\nu F_{\mu\lambda} &
      \left(\ J^P=\frac{5}{2}^+\right) \\ 
   \end{array}
   \right.
   \label{LagBar}
\ee
where $f_{J^P}$ are the coupling constants between
$\rho$, $N$ and the baryon resonance $R$,
$F_{\mu\nu}=\partial_\mu\rho_\nu - \partial_\nu\rho_\mu$ is the $\rho$
field strength tensor, $\psi$ is the nucleon field, $\psi^\mu$ is the
spin 3/2 field and $\psi^{\mu\nu}$ is the spin 5/2
field~\cite{Teodorescu, Rushbrooke}. ${\mathcal {F}}_{\rho NR}$ are
hadronic form factors that take into account the finite size of the
particles that appear in the effective vertexes. These form factors
are taken to be of dipole form
\be
   {\mathcal {F}}_{\rho NR}=\left(\frac{2\Lambda^2 + m^2_R}{2\Lambda^2
   + s}\right)^2\, ,
   \label{dipole}
\ee
where $s$ is the energy squared in the system where the resonance
is at rest and $\Lambda$ is a phenomenological cutoff. All the
interactions in Eqs.~(\ref{LagBar}) are current and parity conserving. 

To compute the one-loop $\rho$ self-energy we use the 
generalized Rarita-Schwinger propagators for fields with spin higher
than 1/2, given by~\cite{Teodorescu, Brudnoy, Napsuciale} [hereafter,
four-momenta are represented by capital letters and their components by
lower case letters]
\begin{widetext}
\be
   {\mathcal {R}}^{\mu\nu}_{3/2}(K)&=&\frac{(K\!\!\!\!\!\!\!
   \not\,\,\,\ +\ m_R)}{K^2 - m_R^2}
   \left\{-g^{\mu\nu} + \frac{1}{3}\gamma^\mu\gamma^\nu
   + \frac{2}{3}\frac{K^\mu K^\nu}{m_R^2} - 
   \frac{1}{3}\frac{K^\mu\gamma^\nu - K^\nu\gamma^\mu}{m_R}
   \right\}\label{firsRS}\\
   {\mathcal {R}}^{\alpha\beta\rho\sigma}_{5/2}(K)&=&
   \sum_{\stackrel{\textstyle{\rho\leftrightarrow\sigma}}
   {\textstyle{\alpha\leftrightarrow\beta}}}
   \left\{\frac{1}{10}\frac{K^\alpha K^\beta K^\rho K^\sigma}{m_R^4}
   + \frac{1}{10}\frac{K^\alpha K^\beta K^\sigma\gamma^\rho - 
   K^\rho K^\sigma K^\alpha\gamma^\beta}{m_R^3} 
   + \frac{1}{10}\frac{K^\alpha\gamma^\beta K^\sigma\gamma^\rho}
   {m_R^2}\right.\nonumber\\ 
   &+&
   \frac{1}{20}\frac{K^\alpha K^\beta g^{\sigma\rho} 
   + K^\sigma K^\rho g^{\alpha\beta}}{m_R^2} - \frac{2}{5}
   \frac{K^\alpha K^\sigma g^{\beta\rho}}{m_R^2}
   - \frac{1}{10}
   \frac{\gamma^\sigma K^\beta g^{\alpha\rho} - 
   \gamma^\alpha K^\sigma g^{\beta\rho}}{m_R}\nonumber\\
   &-&\left.\frac{1}{10}
   \gamma^\alpha\gamma^\sigma g^{\beta\rho} -
   \frac{1}{20}g^{\alpha\beta}g^{\sigma\rho} + 
   \frac{1}{4}g^{\alpha\rho}g^{\beta\sigma}\right\}
   \frac{(K\!\!\!\!\!\!\!
   \not\,\,\,\ +\ m_R)}{K^2 - m_R^2}
   \label{rarita}
\ee
\end{widetext}
where $m_R$ is the mass of
the resonance and the sum over the indexes $\alpha\beta\rho\sigma$ of
a tensor $T^{\alpha\beta\rho\sigma}$ means
\be
   \sum_{\stackrel{\textstyle{\rho\leftrightarrow\sigma}}
   {\textstyle{\alpha\leftrightarrow\beta}}} T^{\alpha\beta\rho\sigma}
   \equiv T^{\alpha\beta\rho\sigma} + T^{\beta\alpha\rho\sigma} 
   + T^{\alpha\beta\sigma\rho} + T^{\beta\alpha\sigma\rho}\, .
   \label{defT}
\ee
In order to ensure that the unphysical spin--1/2 degrees of freedom
contained in $\psi^\mu$ and $\psi^{\mu\nu}$ have no observable effects
even in the interacting theories described by Eqs.~(\ref{LagBar}),
the propagators in Eqs.~(\ref{firsRS}) and~(\ref{rarita}) have to be
regarded as the leading order terms in an expansion in the parameter
$1/m_R$~\cite{Napsuciale}.

We also include the contribution from interactions between $\rho$ and
nucleons given by 
\be
   {\mathcal {L}}_{\mbox{\tiny{$\rho NN$}}}
   =f_{\mbox{\tiny{$\rho NN$}}}
   \bar{\psi}\left(\gamma^\mu -
   \frac{\kappa}{2m_N}\sigma^{\mu\nu}\partial_\nu\right)
   \rho_\mu\psi
   \label{lagrNN}
\ee
where $\sigma^{\mu\nu}=(i/2)[\gamma^\mu,\gamma^\nu]$, $m_N$ is the mass
of the nucleon and we take the values of the dimensionless coupling
constants $f_{\mbox{\tiny{$\rho NN$}}}$ and $\kappa$ 
as $f_{\mbox{\tiny{$\rho NN$}}}=2.63$ and $\kappa=6.1$~\cite{Machleidt}.

\begin{figure}[t!] 
\hspace{-1.3cm}
{\centering
\resizebox*{0.4\textwidth}
{0.18\textheight}{\includegraphics{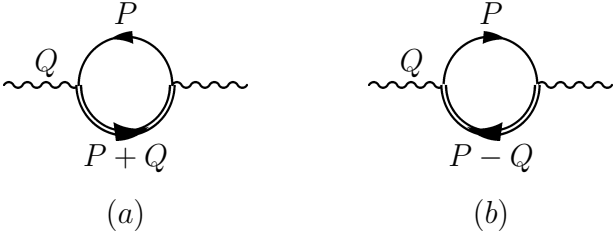}}
\par}
\caption{Feynman diagrams representing the contribution to the $\rho$
one-loop self-energy from interactions with nucleons (single lines) and
baryon resonances (double lines). Both diagrams (a) and (b) need to be
considered as distinct since the particles in the loop are not the
antiparticles of each other.}
\label{fig4}
\end{figure}

For the interaction Lagrangians in Eqs.~(\ref{LagBar}), the
calculation involves the sum of the two Feynman diagrams shown in
Fig.~\ref{fig4} whose corresponding expressions, written in Minkowski space are
\be
   \Pi^{\pm (a)}_{(J)\mu\nu} &\!\!\!=\!\!\!& IF
   \!\!\int\frac{d^4p}{(2\pi )^4}
   \frac{T^{\pm}_{(J)\mu\nu}(P,P+Q)}{[P^2-m_N^2][(P+Q)^2-m_R^2]}
   \nonumber\\
   \Pi^{\pm (b)}_{(J)\mu\nu} &\!\!\!=\!\!\!& IF
   \!\!\int\frac{d^4p}{(2\pi )^4}
   \frac{T^{\pm}_{(J)\mu\nu}(P,P-Q)}{[P^2-m_N^2][(P-Q)^2-m_R^2]}
   \label{loopint}
\ee
where $\pm$ refer to the case of interactions with
positive and negative parity baryon resonances, respectively, $IF$ is
the isospin factor and
\be
   \!\!\!\!\!\!\!\!T^{\pm}_{(3/2)\mu\nu}(P,K)&\!\!\!\!=\!\!\!\!&
   {\mbox{Tr}}[(P\!\!\!\!\!\!\!\not\,\,\,\,
   + m_N)\Gamma^{\pm}_{\mu\alpha}{\mathcal {R}}^{\alpha\beta}_{3/2}(K)
   \Gamma^{\pm}_{\beta\nu}]\nonumber\\
   \!\!\!\!\!\!\!\!T^{\pm}_{(5/2)\mu\nu}(P,K)&\!\!\!\!=\!\!\!\!&
   {\mbox{Tr}}[(P\!\!\!\!\!\!\!\not\,\,\,\,
   + m_N)\Gamma^{\pm}_{\mu\alpha\beta}
   {\mathcal {R}}^{\alpha\beta\rho\sigma}_{5/2}(K)
   \Gamma^{\pm}_{\rho\sigma\nu}]
   \label{tensors}
\ee
where the vertices $\Gamma^{\pm}_{\mu\alpha}$ and
$\Gamma^{\pm}_{\mu\alpha\beta}$, as obtained from the interaction
Lagrangians in Eqs.~(\ref{LagBar}), are given by
\be
   \Gamma^{\pm}_{\mu\alpha}&=&
   \left(\frac{f_{3/2^\pm}}{m_\rho}\right)
   {\mathcal {F}}_{\rho N3/2^\pm}\gamma_5^{(1\pm 1)/2}
   \left(\gamma_\mu Q_\alpha - Q\!\!\!\!\!\!\not\,\,\,\,g_{\mu\alpha}
   \right)\nonumber\\
   \Gamma^{\pm}_{\mu\alpha\beta}&=&
   \left(\frac{f_{5/2^\pm}}{m_\rho}\right)
   {\mathcal {F}}_{\rho N5/2^\pm}\gamma_5^{(1\pm 1)/2}\gamma_\delta Q^\beta
   \nonumber\\
   &&\left(Q_\alpha g_{\delta\mu} - Q_\delta g_{\alpha\mu}\right)\, .
   \label{vertices}
\ee

It is a lengthy though straightforward exercise to compute the
tensors in Eqs.~(\ref{tensors}) with the result
\begin{widetext}
\be
   T^{\pm}_{(3/2)\mu\nu}(P,P+Q)&=&-\frac{8}{3}Q^2
   \left(\frac{f_{3/2^\pm}}{m_\rho}\right)^2
   {\mathcal {F}}_{\rho N3/2^\pm}^2\Big\{
   \left[1+\frac{P\cdot Q +P^2}{m_R^2}\right]
   K_{\mu\nu}\nonumber\\
   &+&
   \left[ P^2\pm m_Nm_R - \frac{(P\cdot Q)^2}{Q^2}
   - \left(\frac{(P^2 + P\cdot Q)(Q^2 + P\cdot Q)^2}{m_R^2Q^2}
   \right)\right]Q_{\mu\nu}\Big\}
   \nonumber\\
   T^{\pm}_{(3/2)\mu\nu}(P,P-Q)&=&-\frac{8}{3}Q^2
   \left(\frac{f_{3/2^\pm}}{m_\rho}\right)^2
   {\mathcal {F}}_{\rho N3/2^\pm}^2\Big\{
   \left[1-\frac{P\cdot Q - P^2 + 4Q^2}{m_R^2}\right]
   K_{\mu\nu}\nonumber\\
   &+&
   \left[ P^2\pm m_Nm_R - \frac{(P\cdot Q)^2}{Q^2}
   - \left(\frac{(P^2 - P\cdot Q)(Q^2 - P\cdot Q)^2}{m_R^2Q^2}
   \right)\right]Q_{\mu\nu}\Big\}
   \nonumber\\
   T^{+}_{(5/2)\mu\nu}(P,P+Q)&=&-\frac{8}{5}Q^2
   \left(\frac{f_{5/2^+}}{m_\rho}\right)^2
   {\mathcal {F}}_{\rho N5/2^+}^2\Big\{
   \left[
   \frac{
   2(Q^2 + P\cdot Q)^2 + 2(Q^2 + P\cdot Q)Q^2 - 3(P^2 + P\cdot Q)Q^2}
   {4m_R^2}\right.
   \nonumber\\
   &-&\left.\left(1-\frac{m_N}{4m_R}\right)Q^2 + \frac{(Q^2 + P\cdot
   Q)^2(P^2 + P\cdot Q)}{m_R^4}\right]K_{\mu\nu}
   + \left[
   \left(\frac{6m_N}{8m_R}\right)(Q^2 + P\cdot Q)^2
   \right.\nonumber\\
   &-&\frac{(Q^2 + P\cdot Q)^4(P^2 + P\cdot Q)}{m_R^4Q^2}
   - \left(\frac{Q^2}{4}\right) 
   \left(3P^2 + 3m_Nm_R - \frac{2(P\cdot Q)^2}{Q^2} +
   P\cdot Q\right)\nonumber\\
   &+&\left.
   \frac{7(Q^2 + P\cdot Q)^2(P^2 + P\cdot Q)Q^2
   - 2(P^2 + P\cdot Q)^3
   (P\cdot Q)}{4m_R^2Q^2}\right]Q_{\mu\nu}\Big\}
   \nonumber\\
   T^{+}_{(5/2)\mu\nu}(P,P-Q)&=&-\frac{8}{5}Q^2
   \left(\frac{f_{5/2^+}}{m_\rho}\right)^2
   {\mathcal {F}}_{\rho N5/2^+}^2\Big\{
   \left[
   \frac{
   2(Q^2 - P\cdot Q)^2 - 2(Q^2 - P\cdot Q)Q^2 - 3(P^2 - P\cdot Q)Q^2}
   {4m_R^2}\right.
   \nonumber\\
   &-&\left.\left(1-\frac{m_N}{4m_R}\right)Q^2 + \frac{(Q^2 - P\cdot
   Q)^2(P^2 - P\cdot Q)}{m_R^4}\right]K_{\mu\nu}
   + \left[
   \left(\frac{6m_N}{8m_R}\right)(Q^2 - P\cdot Q)^2
   \right.\nonumber\\
   &-&\frac{(Q^2 - P\cdot Q)^4(P^2 - P\cdot Q)}{m_R^4Q^2}
   - \left(\frac{Q^2}{4}\right) 
   \left(3P^2 + 3m_Nm_R - \frac{2(P\cdot Q)^2}{Q^2} -
   P\cdot Q\right)\nonumber\\
   &+&\left.
   \frac{7(Q^2 - P\cdot Q)^2(P^2 - P\cdot Q)Q^2
   - 2(P^2 - P\cdot Q)^3
   (P\cdot Q)}{4m_R^2Q^2}\right]Q_{\mu\nu}\Big\}
   \label{tens2}
\ee
\end{widetext}
where
\be
   Q_{\mu\nu}&=&\left(-g_{\mu\nu} +\frac{Q_\mu Q_\nu}{Q^2}\right)
   \nonumber\\
   K_{\mu\nu}&=&
   \left(P_\mu - \frac{(P\cdot Q)}{Q^2}Q_\mu\right)
   \left(P_\nu - \frac{(P\cdot Q)}{Q^2}Q_\nu\right)
   \label{structures}
\ee   
and for the last two of Eqs.~(\ref{tens2}) we have only considered the
case of baryon resonances with positive parity interacting with
$\rho$. The fact that Eqs.~(\ref{tens2}) can be written in terms of
the tensor structures $Q_{\mu\nu}$ and $K_{\mu\nu}$ makes it evident
that these are explicitly transverse.

The nucleon isospin factor is taken as $IF=2$, both for the case of
interactions with N or $\Delta$ resonances as regards to the
calculation of mass modifications of $\rho^0$. To see why this is so,
we must keep in mind that the real part of the self-energy amounts to
the sum of the squares of amplitudes that represent scattering
processes where $\rho^0$ takes part. Consider for instance the
$s$-channel scattering of $\rho^0$ and a nucleon with the exchange of
an intermediate N or $\Delta$ resonance. Since there is conservation
of charge at the vertex, the number of possibilities for this
scattering to take place is fixed by the number of nucleon states in
the initial or final state which is equal to 2, regardless of whether
the exchanged particle is an N or a $\Delta$ resonance. 

\subsection{Coupling constants}

Before going into the computation of the thermal modification to the
$\rho$ mass, let us pause for a moment and compute the coupling
constants $f_{J^P}$ appearing in Eqs.~(\ref{tens2}). This is
accomplished by comparing the theoretical expression for the branching
ratio of the given baryon resonance into the $\rho-N$ channel
with the experimentally measured value. 

\begin{figure}[t!] 
\hspace{-1.3cm}
{\centering
\resizebox*{0.18\textwidth}
{0.135\textheight}{\includegraphics{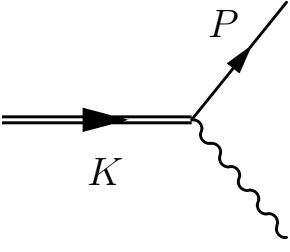}}
\par}
\caption{Feynman diagram representing the amplitude for a given
baryon resonance (double line) to decay into the $\rho$ (wavy line)
and nucleon (single line) channel.}
\label{fig5}
\end{figure}

Figure~\ref{fig5} represents the amplitude for a given baryon
resonance to decay into the $\rho-N$ channel, where the kinematics is
also defined. The expression for the partial width for such process in
the rest frame of the decaying resonance is given by 
\be
   \Gamma_{R\rightarrow\rho N}=\frac{p}{32\pi^2 m_R^2}\int
   d\Omega |M_{R\rightarrow\rho N}|^2\, ,
   \label{width}
\ee
where $|M_{R\rightarrow\rho N}|^2$ represents the matrix element
squared summed over final and averaged over initial spin and isospin
states. For the interactions considered in Eqs.~(\ref{LagBar}) it is
easy to see that these can be written as
\be
   |M_{3/2^\pm\rightarrow\rho N}|^2&\!\!\!=\!\!\!&
   \frac{IF\ (2I_{R,\rho}+1)}{(2J_R+1)(2I_R+1)}
   T^{\pm}_{(3/2)\alpha\beta}(P,K)
   \nonumber\\
   &\!\!\!\times\!\!\!&                 
   \left(-g^{\alpha\beta}
   +\frac{Q^\alpha Q^\beta}{Q^2}\right)
   \nonumber\\
   |M_{5/2^+\rightarrow\rho N}|^2&\!\!\!=\!\!\!&\frac{IF\
   (2I_\rho+1)}{(2J_R+1)(2I_R+1)}
   T^{+}_{(5/2)\alpha\beta}(P,K)
   \nonumber\\ 
   &\!\!\!\times\!\!\!&
   \left(-g^{\alpha\beta}
   +\frac{Q^\alpha Q^\beta}{Q^2}\right)
   \label{matsqua}
\ee
where $Q=K-P$ and in the first of Eqs.~(\ref{matsqua}) the factor
$(2I_R+1)$ describes the decay of N resonances and the factor
$(2I_\rho+1)$ the case of the $\Delta$ resonance. 

In the rest frame of the decaying resonance ${\mathbf {k}}=0$ and
$P^2=m_N^2$, $P\cdot K=E_Nm_R$, thus, it is straightforward to show
that $|M_{R\rightarrow\rho N}|^2$ are given by
\begin{widetext}
\be
   |M_{3/2^\pm\rightarrow\rho N}|^2&=&
   \frac{IF\ (2I_{R,\rho}+1)}{(2J_R+1)(2I_R+1)}
   \left(-\frac{8}{3}Q^2\right)
   \left(\frac{f_{3/2^\pm}}{m_\rho}\right)^2
   {\mathcal {F}}_{\rho N3/2^\pm}^2
   \Big\{\left[m_N^2 - \frac{(E_Nm_R - m_N^2)^2}{Q^2}\right]
   \left[ 2 + \frac{(4Q^2 + m_N^2)}{m_R^2}\right]
   \nonumber\\
   &+&3\left[\pm m_Nm_R - \frac{(E_Nm_R)}{m_R^2Q^2}
   (Q^2 + E_Nm_R -m_N^2)^2\right]\Big\}
   \nonumber\\
   |M_{5/2^+\rightarrow\rho N}|^2&=&
   \frac{IF\ (2I_\rho+1)}{(2J_R+1)(2I_R+1)}
   \left(-\frac{8}{5}Q^2\right)
   \left(\frac{f_{5/2^+}}{m_\rho^2}\right)^2
   {\mathcal {F}}_{\rho N5/2^+}^2
   \Big\{\left(\frac{(m_N^2 - E_Nm_R - Q^2)^2
   (E_Nm_R)}{m_R^4}\right)
   \nonumber\\
   &\times&
   \left[\frac{2(E_Nm_R - m_N^2)^2}{Q^2} + 3Q^2 + 6E_Nm_R -
   5m_N^2\right]
   \nonumber\\
   &+&
   \left(\frac{Q^2}{4}\right)\left[\frac{2(E_Nm_R - m_N^2)^2}{Q^2} 
   - 2m_N^2 - 9m_Nm_R - 3E_Nm_R\right]
   + \left(\frac{m_N}{4m_R}\right)
   \left[10(E_Nm_R - m_N^2)^2 + 9Q^4\right. 
   \nonumber\\
   &+&\left.18Q^2E_Nm_R - 19m_N^2Q^2\right]
   -\left(\frac{1}{4m_R^2}\right)
   \left[\frac{4(m_N^2 - E_Nm_R)^4}{Q^2} - 2(m_N^2 - E_Nm_R)^2
   (5m_N^2 + 3E_Nm_R)
   \right.
   \nonumber\\
   &+&\left.
   (8m_N^4 + 17m_N^2E_Nm_R - 28(E_Nm_R)^2)Q^2
   - (2m_N^2 + 15E_Nm_R)Q^4\right]\Big\}\, .
   \label{matexpli}
\ee
\end{widetext}   

For a reliable estimate of the coupling constants, we should include
the finite width of $\rho$ by folding the expression for the width,
Eq.~(\ref{width}), at a given value $Q^2$ of the $\rho$ mass with the
$\rho$ spectral function $S(Q)$. 
\be
   \Gamma_{R\rightarrow\rho N}^{\mbox{\tiny{finite width}}}=
   \int_{2m_\pi}^{m_R-m_N}\frac{QdQ}{\pi}S(Q)
   \Gamma_{R\rightarrow\rho N}\, .
   \label{gamafinite}
\ee
For definitiveness, $S(Q)$ is taken as a relativistic
Breit--Wigner function  
\be
   S(Q)=\frac{2m_\rho\Gamma (Q^2)}{(Q^2-m_\rho^2)^2 + 
   (m_\rho\Gamma(Q^2))^2}\, , 
   \label{BW}
\ee
where we also include the proper phase space angular momentum
dependence for the $\rho$ decay into two
pions~\cite{Peters}, taking  
\be
   \Gamma (Q^2)=\Gamma^{\mbox{\tiny{vac}}}\left(
   \frac{\sqrt{Q^2/4-m_\pi^2}}{\sqrt{m_\rho^2/4-m_\pi^2}}\right)^3\, ,
\ee
where we use $\Gamma^{\mbox{\tiny{vac}}}=150$ MeV.

Table~I shows the values of the coupling constants
$f_{J^P}$ obtained by equating Eq.~(\ref{gamafinite}) to the
experimentally measured values of
$\Gamma^{\mbox{\tiny{vac}}}_{R\rightarrow\rho N}$ also listed in the
table. For comparison, we 
also show the values of the coupling constants
$f_{J^P}^{\mbox{\tiny{N--R}}}$ obtained from a non-relativistic
approach~\cite{Peters}. In the calculations we set $\Lambda=1$
GeV, however, the obtained values for the coupling constants do not
change when $\Lambda$ varies in the range $1$ GeV $<\Lambda<$ $2$ GeV
which is a reasonable interval when considering hadronic processes. 

\subsection{Thermal mass}

We now proceed to the computation of the thermal modifications of the
$\rho$ mass. For this purpose, we work in the imaginary-time
formulation of thermal field theory. Thus, the corresponding
expressions for the $\rho$ self-energy of Eqs.~(\ref{loopint}) become
\be
   \Pi^{\pm (a)}_{(J)\mu\nu} &=& T\sum_n\int\frac{d^3p}{(2\pi )^3}
   \frac{T^{\pm}_{(J)\mu\nu}(P,P+Q)}{[P^2-m_N^2][(P+Q)^2-m_R^2]}
   \nonumber\\
   \Pi^{\pm (b)}_{(J)\mu\nu} &=& T\sum_n\int\frac{d^3p}{(2\pi )^3}
   \frac{T^{\pm}_{(J)\mu\nu}(P,P-Q)}{[P^2-m_N^2][(P-Q)^2-m_R^2]}\, ,
   \nonumber\\
   \label{loopint2}
\ee
where $p_0=i\omega_n$ and $q_0=i\omega$, with $\omega_n=(2n+1)\pi T$
and $\omega=(2m+1)\pi T$ being fermion Matsubara
frequencies and $n$, $m$ integers.

For definitiveness, we take the $\hat{z}$ axis as
the direction of motion of the $\rho$ meson and thus the square of its
thermal mass can be computed from the thermal part of the component
$\Pi_{(J)11}^{\pm}=\Pi_{(J)11}^{\pm (a)}+\Pi_{(J)\ 11}^{\pm (b)}$
of the $\rho$ self-energy, in the limit of vanishing 
three-momentum~\cite{Gale}.  

From Eqs.~(\ref{tens2}) and~(\ref{loopint2}), we must carry
out sums over Matsubara frequencies of the form
\be
   \tilde{S}_l=T\sum_n(i\omega_n)^l\tilde{\Delta}(i\omega_n,E_N)
   \tilde{\Delta}(i(\omega-\omega_n),E_R)\, ,
   \label{sums}
\ee
where $l=0,\ldots , 7$ and
\be
   \tilde{\Delta}(i\omega_n,E)=\frac{1}{\omega_n^2+E^2}\, .
   \label{tildeD}
\ee
The sums for the cases $l=0,1$ are well known~\cite{LeBellac} and are
given by
\be
   \tilde{S}_0&=&\sum_{s_1,s_2=\pm}-\frac{s_1s_2}{4E_NE_R}
   \frac{1-\tilde{n}(s_1E_N)-\tilde{n}(s_2E_R)}{i\omega-s_1E_N-s_2E_R}
   \nonumber\\
   \tilde{S}_1&=&\sum_{s_1,s_2=\pm}-\frac{s_2}{4E_R}
   \frac{1-\tilde{n}(s_1E_N)-\tilde{n}(s_2E_R)}{i\omega-s_1E_N-s_2E_R}\, ,
   \label{sumexp1}
\ee
where
\be
   \tilde{n}(E)=\frac{1}{e^{\beta E}+1}\, ,
   \label{FD}
\ee
is the Fermi-Dirac distribution. In order to carry out the rest of the
sums, we make use of the identity
\be
   \omega_n^l\tilde{\Delta}(i\omega_n,E_N)=\omega_n^{l-2}
   \left[1 - \frac{E_N^2}{\omega_n^2+E_N^2}\right]\, ,
   \label{identity}
\ee
for $l\geq 2$, together with the result
\be
   &&T\sum_n\omega_n^l\tilde{\Delta}(i\omega_n,E_R)\rightarrow
   \nonumber\\
   && \left\{\begin{array}{cr}
              -(i)^lE_R^{l-1}\tilde{n}(E_R) & l\ {\mbox {even or}}\ 0\\
                             0              & l\ {\mbox {odd}}
          \end{array}\right.\, ,
   \label{identity2}
\ee
where the arrow indicates that we just consider the temperature
dependent terms. Using Eqs.~(\ref{identity}) and~(\ref{identity2}) it
is easy to show that 
\be
   \tilde{S}_2&\rightarrow&E_N^2\tilde{S}_0 +
   \frac{\tilde{n}(E_R)}{E_R}\nonumber\\
   \tilde{S}_3&\rightarrow&E_N^2\tilde{S}_1 + (i\omega)
   \frac{\tilde{n}(E_R)}{E_R}\nonumber\\
   \tilde{S}_4&\rightarrow&E_N^4\tilde{S}_0 + (E_R^2+E_N^2-\omega^2)
   \frac{\tilde{n}(E_R)}{E_R}\nonumber\\
   \tilde{S}_5&\rightarrow&E_N^4\tilde{S}_1 + (i\omega)(3E_R^2+E_N^2-\omega^2)
   \frac{\tilde{n}(E_R)}{E_R}\nonumber\\
   \tilde{S}_6&\rightarrow&E_N^6\tilde{S}_0 + (-6\omega^2E_R^2+E_R^4+\omega^4
   \nonumber\\
   &+&E_N^2E_R^2-E_N^2\omega^2+E_N^4)
   \frac{\tilde{n}(E_R)}{E_R}\nonumber\\
   \tilde{S}_7&\rightarrow&E_N^6\tilde{S}_1 + (i\omega)
   (\omega^4-10\omega^2E_R^2+5E_R^4
   \nonumber\\
   &+&3E_N^2E_R^2+E_N^4-E_N^2\omega^2)
   \frac{\tilde{n}(E_R)}{E_R}\, .
   \label{allsums}
\ee
Using Eqs.~(\ref{sumexp1}) and~(\ref{allsums}) into
Eqs.~(\ref{loopint2}) we can now 
make the analytical continuation $i\omega\rightarrow q_0 + i\epsilon$
to compute the real part of the self-energies $\Pi_{(J)11}^{\pm}$, in
a similar fashion as the one explained in detail in Sec.~\ref{secII}. 

\vspace{0.9cm}
\begin{figure}[t] 
\vspace{0.4cm}
{\centering
\resizebox*{0.4\textwidth}
{0.2\textheight}{\includegraphics{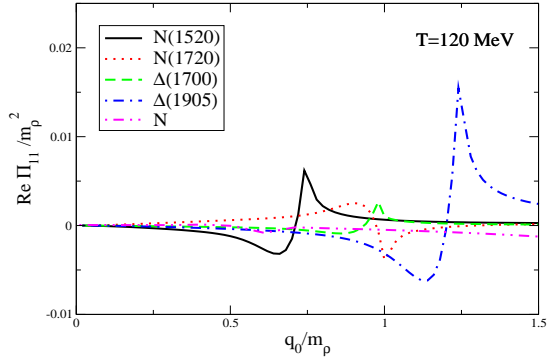}}
\par}
\caption{Contribution to the $\rho$ self-energy from scattering with
  nucleons and various other nucleon resonances. The main contribution
  for $q_0\sim m_\rho$ stems  from the formation of $s$-channel,
  positive parity spin 3/2 N(1720) and spin 5/2 $\Delta$(1905)
  resonances.} 
\label{fig6}
\end{figure}

The result is summarized in Fig.~\ref{fig6} where we show the
temperature dependent real part of the self-energies scaled to the
square of the $\rho$ mass in vacuum, for each 
of the resonances listed in Table~I as a function 
of $q_0/m_\rho$, where $q_0$ is the energy of the $\rho$ meson at
rest, for a temperature $T=120$ MeV. Notice that the
two Feynman diagrams of Fig.~\ref{fig5} represent the contribution to
the scattering process from nucleons and anti-nucleons, as corresponds to
the scenario where resonance production happens from the --almost
baryon free-- central region of the reaction. The main contributions
in magnitude for $q_0\sim  m_\rho$ come from the resonances with even
parity N(1720) and $\Delta$(1905). We also show the contribution from
scattering with nucleons, taking $m_N$ to its vacuum value. Notice
that for the kinematical range considered, the contribution from
nucleons is completely negligible. 

\section{Interactions of $\rho$ with pions and other mesons}\label{secIV}

We now look at the contribution to the $\rho$ self-energy stemming from
scattering with pions and other mesons. Table~II shows the
quantum numbers of those mesons with sizable branching ratios
involving the pion and $\rho$ and whose rest mass is near the sum of
the rest masses of the pion/kaon and $\rho$, as well as the coupling
constants used in the calculation. For the interaction of $\rho$ and
the pion, $\pi$, we take the Lagrangian
\be
   {\mathcal {L}} = g_{\mbox{\tiny{$\rho\pi\pi$}}}
                    \pi(\stackrel{\leftrightarrow}{\partial}\ \!\!\! ^\mu
                     - ig_{\mbox{\tiny{$\rho\pi\pi$}}}\rho^\mu)\pi
                    \rho_\mu
   \label{lagpipirho}
\ee
obtained by {\it gauging} the pion-pion interaction
Lagrangian~\cite{Gale, Ayala}. The value of the dimensionless coupling
constant, as determined by the $\rho$ width in vacuum, is taken as
$g_{\mbox{\tiny{$\rho\pi\pi$}}}=6.06$~\cite{Gale}. 

The interaction Lagrangians involving the other relevant mesons are
taken as~\cite{Rapp}  
\be
   {\mathcal {L}} = \left\{ 
   \begin{array}{l}
   g_{\mbox{\tiny{$\rho PA$}}}{\mathcal {F}}_{\rho PA}
   \ A^\mu\partial^\nu P F_{\mu\nu} \\
   g_{\mbox{\tiny{$\rho\pi V$}}}{\mathcal {F}}_{\rho\pi V}
   \ \epsilon_{\mu\nu\sigma\tau}
   \partial^\mu \pi V^\nu\partial^\sigma\rho^\tau\\
   (g_{\mbox{\tiny{$\rho\pi P'$}}}/m_\rho){\mathcal {F}}_{\rho\pi P'} 
   \ \partial^\mu \pi\partial^\nu P'F_{\mu\nu}\\ 
   \end{array}
   \right.
   \label{lagothermesons}
\ee 
where $V^\mu$ and $A^\mu$ represent the vector and axial-vector fields
and $P$, $P'$ the pseudo-scalar fields. ${\mathcal {F}}_{\rho PR}$ 
are dipole form factors [see Eq.~(\ref{dipole})]. The coupling
constants $g_{\mbox{\tiny{$\rho PR$}}}$ in
Eqs.~(\ref{lagothermesons}) are taken from 
Ref.~\cite{Rapp}. The interaction Lagrangians in 
Eqs.~(\ref{lagpipirho}) and~(\ref{lagothermesons}) are current and
parity conserving as well as compatible with chiral symmetry.

\begin{table}[t!]
\begin{tabular}{|ccccccc|}  \hline\hline
\multicolumn{1}{|c}{$R$}  &
\multicolumn{1}{c}{$J^P$} &
\multicolumn{1}{c}{$\rho h$ decay} &
\multicolumn{1}{c}{$\Gamma^{\mbox{\tiny{vac}}}_{\rho h}$} &
\multicolumn{1}{c}{$\Gamma^{\mbox{\tiny{vac}}}_{\mbox{\tiny{tot}}}$} &
\multicolumn{1}{c}{$g_{\rho hR}$} &
\multicolumn{1}{c|}{$IF$} \\ 
& & & ${\mbox{(MeV)}}$ & ${\mbox{(MeV)}}$ & ${\mbox{(GeV)}}^{-1}$ &
\\ \hline
$\omega$(782) & $1^-$ & $\rho\pi$ & $\sim 5$ & $8.43$ & 25.8 & 1 \\
$h_1$(1170)   & $1^+$ & $\rho\pi$ &   seen   & $\sim 360$ & 11.37
& 1 \\
$a_1$(1260)   & $1^+$ & $\rho\pi$ & dominant & $\sim 400$ & 13.27 
& 2 \\ 
$K_1$(1270)   & $1^+$ & $\rho K$  & $\sim 60$& $\sim 90$  & 9.42
& 2 \\
$\pi'$(1300)  & $0^-$ & $\rho\pi$ &   seen   & $\sim 400$ & 7.44
& 2 \\ \hline
\end{tabular}
\label{tab2}
\caption{List of meson resonances included in the calculation. The
  last column corresponds to the isospin factor accounting for the
  number of isospin channels that take part in the dispersion. The
  coupling constants are taken from the analysis in Ref.~\cite{Rapp}.}
\vspace{-0.5cm}
\end{table}

The expressions for the one-loop $\rho$ self-energy corresponding
to the interaction Lagrangians in Eqs.~(\ref{lagothermesons}), can be
written in Minkowski space, as 
\be
   \Pi^{(J^P)}_{\mu\nu} &\!\!\!=\!\!\!& IF\int\frac{d^4p}{(2\pi )^4}
   \frac{M^{(J^P)}_{\mu\nu}}{[P^2-m_{\mbox{\tiny{P}}}^2][(P-Q)^2-m_R^2]}
   \label{loopintmesons}
\ee
where $m_{\mbox{\tiny{P}}}$ is the mass of the pseudoscalar ($\pi$ or
$K$) and $m_R$ is the mass of the vector, axial-vector or
$\pi'(1300)$ and $IF$ is the isospin factor. The numerators in
Eq.~(\ref{loopintmesons}) are given by
\be
   M^{(1^\pm)}_{\mu\nu}&=&\Gamma_{\mu\alpha}^{(1^\pm)}
   \left(g^{\alpha\beta} - \frac{K^\alpha K^\beta}{m_{R}^2}\right)
   \Gamma_{\beta\nu}^{(1^\pm)}\nonumber\\
   M^{(0^+)}_{\mu\nu}&=&\Gamma_\mu^{(0^+)}\Gamma_\nu^{(0^+)}\, ,
   \label{Mmesons}
\ee
where $K=Q-P$ and the vertices, as obtained from the interaction
Lagrangians in Eqs.~(\ref{lagothermesons}), are given by 
\be
   \Gamma_{\alpha\beta}^{(1^+)}&=&g_{\mbox{\tiny{$\rho PA$}}}
   {\mathcal{F}}_{\rho PA}\
   [g_{\alpha\beta}(P\cdot Q) - P_\alpha Q_\beta]\nonumber\\
   \Gamma_{\alpha\beta}^{(1^-)}&=&g_{\mbox{\tiny{$\rho\pi V$}}} 
   {\mathcal {F}}_{\rho\pi V}\ 
   [\epsilon_{\gamma\alpha\delta\beta}(Q-P)^\gamma Q^\delta]\nonumber\\
   \Gamma_{\alpha}^{(0^+)}&=&(g_{\mbox{\tiny{$\rho\pi P'$}}}/m_\rho) 
   {\mathcal {F}}_{\rho\pi P'}\nonumber\\
   &&[Q\cdot (Q-P)P_\alpha - (P\cdot Q)(Q-P)_\alpha]
   \label{verticesmesons}
\ee
It is easy to check that the explicit expressions for
$M^{(1^\pm)}_{\mu\nu}$ and $M^{(0^+)}_{\mu\nu}$ are given by
\be
   M^{(1^+)}_{\mu\nu}&=&g_{\mbox{\tiny{$\rho PA$}}}^2
   {\mathcal{F}}_{\rho PA}^2\nonumber\\
   &&\left\{
   [P^2Q^2-(P\cdot Q)^2]\ Q_{\mu\nu} +
   Q^2\ K_{\mu\nu}\right\}\nonumber\\
   M^{(1^-)}_{\mu\nu}&=&g_{\mbox{\tiny{$\rho\pi V$}}}^2
   {\mathcal {F}}_{\rho\pi V}^2\nonumber\\
   &&\left\{-(P\cdot Q)^2\ Q_{\mu\nu} +
   Q^2\left(1-\frac{Q^2}{m_R^2}\right)K_{\mu\nu}\right\}\nonumber\\
   M^{(0^+)}_{\mu\nu}&=&(g_{\mbox{\tiny{$\rho\pi P'$}}}/m_\rho)^2
   {\mathcal {F}}_{\rho\pi P'}^2
   Q^4\ K_{\mu\nu}\, ,
   \label{Mexp}
\ee
where the presence of the tensors $Q_{\mu\nu}$ and $K_{\mu\nu}$,
defined in Eqs.~(\ref{structures}), makes it evident that
Eqs.~(\ref{Mexp}) are explicitly transverse.

For the computation of the contributions to the thermal modifications
of the $\rho$ mass, we work in the imaginary-time formalism of thermal
field theory writing Eq.~(\ref{loopintmesons}) in Euclidian space
\be
   \!\!\!\!\!\!
   \Pi^{(J^P)}_{\mu\nu} &\!\!\!=\!\!\!& T\sum_n\int\frac{d^3p}{(2\pi )^3}
   \frac{M^{(J^P)}_{\mu\nu}}{[P^2-m_{\mbox{\tiny{P}}}^2][(Q-P)^2-m_R^2]}
   \label{loopintmesonsEuc}
\ee
where $p_0=i\omega_n$ and $q_0=i\omega$, with $\omega_n=2n\pi T$ and
$\omega=2m\pi T$ being boson Matsubara frequencies, and $n$ and $m$
integers. Once again, for definitiveness, we take the $\hat{z}$ axis as 
the direction of motion of the $\rho$ meson and thus the square of its
thermal mass can be computed from the thermal part of the component
$\Pi^{(J^P)}_{11}$ in Eq.~(\ref{loopintmesons}) in the limit of
vanishing three-momentum.

From Eqs.~(\ref{Mexp}) and~(\ref{loopintmesonsEuc}), we must carry out
sums over Matsubara frequencies of the form
\be
   S_l=T\sum_n(i\omega_n)^l\Delta(i\omega_n,E_N)
   \Delta(i(\omega-\omega_n),E_R)\, ,
   \label{sumsbos}
\ee
where $l=0,1,2$ and
\be
   \Delta(i\omega_n,E)=\frac{1}{\omega_n^2+E^2}\, .
   \label{notildeD}
\ee
The sums for the cases $l=0,1$ are well known~\cite{LeBellac} and are
given by
\be
   S_0&=&\sum_{s_1,s_2=\pm}-\frac{s_1s_2}{4E_PE_R}
   \frac{1+n(s_1E_P)+n(s_2E_R)}{i\omega-s_1E_P-s_2E_R}
   \nonumber\\
   S_1&=&\sum_{s_1,s_2=\pm}-\frac{s_2}{4E_R}
   \frac{1+n(s_1E_P)+n(s_2E_R)}{i\omega-s_1E_P-s_2E_R}\, ,
   \label{sumexpbos}
\ee
where $n(E)$ is the Bose-Einstein distribution given in
Eq.~(\ref{BE}).

It is easy to check that the expression for $S_2$ is given by
\be
   S_2\rightarrow E_P^2S_0 -
   \frac{n(E_R)}{E_R}\, ,
   \label{sum2}
\ee
where the arrow indicates that we just consider the temperature
dependent terms.

\begin{figure}[t!] 
{\centering
\resizebox*{0.4\textwidth}
{0.2\textheight}{\includegraphics{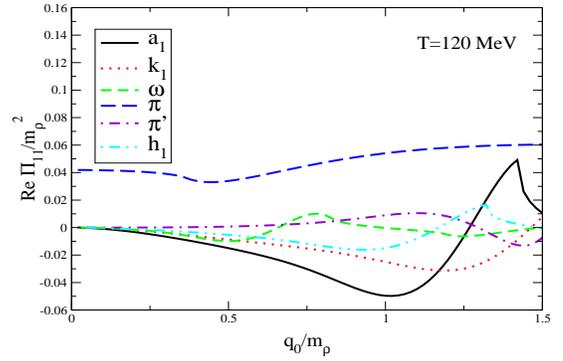}}
\par}
\caption{Contribution to the $\rho$ self-energy from scattering with
  pions and various other mesons. The main contribution for $q_0\sim
  m_\rho$ stems from the formation of an $s$-channel axial-vector
  resonance $a_1$.} 
\label{fig7}
\end{figure}

Using Eqs.~(\ref{sumexpbos}) and~(\ref{sum2}) into
Eq.~(\ref{loopintmesonsEuc}) we can now make the analytical
continuation $i\omega\rightarrow q_0 + i\epsilon$ 
to compute the real part of the self-energies $\Pi_{11}^{(J^P)}$, in
a similar fashion as the one explained in detail in Sec.~\ref{secII}. 
The result is summarized in Fig.~\ref{fig7} where we show the
temperature dependent real part of $\Pi_{11}$ scaled to the square of
the $\rho$ mass in vacuum arising from pion exchange as well as each
of the mesons listed in Table~II as a function of $q_0/m_\rho$, for a
temperature $T=120$ MeV. We notice that in the interval considered,
the main contribution comes from scattering of $\rho$ off
pions. However, for $q_0\sim m_\rho$ a sizable contribution in 
magnitude comes from $\pi$--$\rho$ scattering through the formation of
an $s$-channel axial-vector resonance $a_1$, which has the opposite
sign and about the same strength as the contribution from pion exchange, in
agreement with the findings in Ref.~\cite{Rapp}. Also, for $q_0\sim
m_\rho$, the rest of the contributions offset among themselves.

\section{Intrinsic thermal mass of the $\rho^0$}\label{secV}

\begin{figure}[t!] 
\vspace{0.4cm}
{\centering
\resizebox*{0.4\textwidth}
{0.2\textheight}{\includegraphics{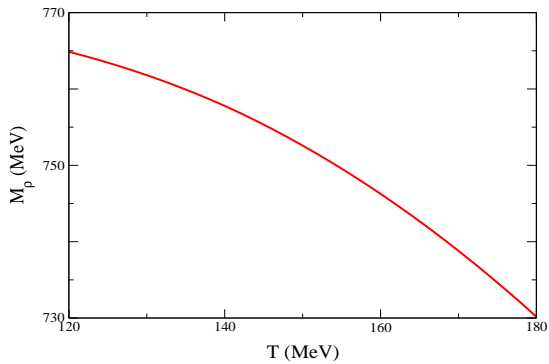}}
\par}
\caption{Intrinsic mass of $\rho^0$ as a
  function of temperature.}
\label{fig8}
\end{figure}

We now put together the contributions from all of the particles
considered in Secs.~III and~IV. This is summarized in Fig.~\ref{fig8}
where we show the total shift in the intrinsic $\rho$ mass as a
function of temperature. Notice that the 
shift is negative and increases in magnitude as the temperature
increases. For instance, taking $m_\rho=770$ MeV, we get
$M_\rho=764$--$730$ MeV when the temperature 
varies between $T=120$--$180$ MeV, which is a reasonable range for the
temperature of the hadronic phase of a relativistic heavy-ion
collision between chemical and kinetic freeze-out.

\section{Summary and conclusions}\label{secVI}

In this work we have computed the intrinsic changes in the $\rho$ mass
due to scattering with the relevant mesons and baryons in the context
of ultra-relativistic heavy-ion collisions, at finite temperature. We
have shown that a consistent field theoretical calculation of the real
part of the one-loop $\rho$ self-energy yields the contributions form
$t$- and $s$-channel scattering and that in general, both have to
be accounted for. In addition to the already well know contributions
arising from $\rho$ scattering with pions through the exchange of
$a_1$ and pions themselves, we have found that the contributions from
scattering with nucleons through the formation of even parity, spin
3/2 [N(1720)] and 5/2 [$\Delta$(1905)] nucleon resonances are significant.

The reason for the difference in the behavior between the real parts
of the contributions of N(1720) and $\Delta (1905)$ to the $\rho$
self-energy as a function of $q_0$ --the former starting out repulsive
and the latter attractive-- is that the structure of their propagators
and couplings with nucleons is different, given that these are
resonances with different spin. The leading term for each case when
$q_0 \rightarrow 0$ is of the form $c*q_0$ where $c$ is a numerical
coefficient to which several terms from the product of the propagator
and vertices contribute. It turns out that this coefficient is
positive in the case of N(1720) and negative in the case of $\Delta
(1905)$. We emphasize that this conclusion is born out of the
explicit calculation. We should however point out that an important
cross check of the result, namely, the transversality of the
self-energy, has been carried out. This is by no means a trivial check
of the consistency of the calculation since had one or more of the
terms that make up the above mentioned $c$ coefficient been wrong, the
transversality would have been spoiled.

The results underline the importance of scattering of $\rho$ mesons
with baryons, in particular nucleons at finite temperature, for the
decrease of the intrinsic mass of $\rho$, during the hadronic phase of
the reaction. 

We have shown that it is possible to achieve a shift in
the intrinsic $\rho^0$ mass of order $\sim - 40$ MeV, when
including the contributions of all the relevant mesons and baryons
that take part in the scattering, for temperatures within the commonly
accepted values for the hadronic phase of the collision, between
chemical and kinetic freeze-out. 
These findings refer to the {\em intrinsic}
changes in the $\rho$ mass. The overall change in the value of the
peak of the invariant $\pi^+\ \pi^-$ distribution should contain
also the effects of phase-space distortions due to thermal motion of
the decay products as well as the effect due to the change in the
intrinsic $\rho$ width~\cite{Kolb, Rapp2}. This is work in progress
and will be reported elsewhere.  

\section*{Acknowledgments}

Support for this work has been received in part by DGAPA-UNAM under
PAPIIT grant number IN108001 and CONACyT under grant number 40025-F.

\end{document}